\documentclass[conference]{IEEEtran}
\IEEEoverridecommandlockouts
\usepackage{cite}
\usepackage{amsmath,amssymb,amsfonts}
\usepackage{algorithmic}
\usepackage{graphicx}
\usepackage{textcomp}
\usepackage{xcolor}
\graphicspath{ {images/} }

\def\BibTeX{{\rm B\kern-.05em{\sc i\kern-.025em b}\kern-.08em
    T\kern-.1667em\lower.7ex\hbox{E}\kern-.125emX}}
\begin{document}

\title{How Camera Placement Affects Gameplay in Video Games\\
\thanks{This research has been co-financed by the European
Union and Greek national funds through the Operational
Program Competitiveness, Entrepreneurship and Innovation,
under the call RESEARCH – CREATE – INNOVATE (project
Mediludus, code: T2EDK-03049)
}}

\author{\IEEEauthorblockN{Markos Naftis}
\IEEEauthorblockA{\textit{Department of Games Programming} \\
\textit{SAE Creative Media College}\\
Moschato, Greece \\
12499gr@saeinstitute.edu}
\and
\IEEEauthorblockN{Georgios Tsatiris}
\IEEEauthorblockA{
\textit{Artificial Intelligence} \\
\textit{and Learning Systems Laboratory} \\
\textit{National Technical University of Athens}\\
Athens, Greece \\
gtsatiris@image.ntua.gr}
\and
\IEEEauthorblockN{Kostas Karpouzis}
\IEEEauthorblockA{
\textit{Dept of Communication, Media and Culture} \\
\textit{Panteion Univ. of Social \& Political Sciences}\\
Athens, Greece \\
kkarpou@cs.ntua.gr}
}

\maketitle

\begin{abstract}
In video games, players' perception of the game world and related information depends on their or the game designer's choice of a virtual camera model. In this paper, we attempt to answer the research question of whether it is possible to identify which camera model is preferred by, fits and best serves each player depending on where they are in a game world and the kinds of challenges they face. To this end, a special type of video game, combining challenges from different game genres, was designed and developed with Unity; thirty players could choose from four camera models at their disposal, depending on where they were in the game world, and utilize the most suitable one to proceed. Each player’s preference of camera model were collected using the data platform Unity Analytics and then analyzed. The analysis of the results showed that players managed to adapt to the logic and requirements of the game challenges by choosing different cameras for each of them, depending on the spatial requirements and the presence of enemies or platforms they should jump across from.
\end{abstract}


\section{Introduction}
In the field of game design, quite a few works deal with camera placement and projection models with respect to player experience. This is an extremely important association, because of its effect on how players perceive the game environment, the amount and nature of information which becomes available to them and their view of the obstacles and enemies they need to overcome in order to progress in the game \cite{thorn2013cameras}. In addition to this, camera placement and model can be a powerful towards the artistic direction of a digital game (termed \emph{gamatography} \cite{kelly}). Nitsche \cite{nitsche2008video} mentions that the contents of the camera view are defined and restricted by what is essential to progress, resulting in a hybrid between architectural navigation and cinematography: players are immersed into a game world and always carry a camera with them, filming a sequence of never-ending movies. Thus, cameras can be thought of as another narrative instrument, besides level design and game mechanics \cite{caridakis2010multimodal}.

This paper investigates whether it is possible to identify a camera model that is fitting to each player's style and preferences, given a particular part of a game level and an obstacle the player is trying to overcome. To this end, we designed and developed a level of a 3D game with different kinds of challenges (gaps and enemies) and enabled players to choose which of the available camera models they preferred for each challenge. Camera choices and player performance were collected and analyzed to provide answers to our research questions. 

\section{Cameras in video games}
\label{cameras}
According to Adams \cite{adams2014fundamentals}, a camera model can be defined as the viewpoint provided by a virtual camera in the game world, along with the instructions and mechanics necessary for it to perform its narrative purpose. This relates to the characteristics and aesthetics of the players' view of the game world, to the game and level elements players are supposed to focus on and on the expressive behavior of the camera \cite{cowie2011issues}. Effectively, this means that camera models can either be \emph{static} or \emph{dynamic}: static models provide a viewpoint from a particular point in the game world, regardless of any short-term game events, while dynamic models react to events in the game and player behavior, making them harder to design and implement, but more effective in terms of narration and directing. More recently, designers employ Game AI techniques to enable cameras to follow the game action and choose the most convenient and expressive viewpoint automatically \cite{jhala2006darshak}, also taking into account the objectives of the player in the particular game (\cite{xu2014generative}, \cite{asteriadis2012towards}).

The most widely used camera models in 3D games are \emph{First-Person} and \emph{Third Person Perspective} cameras, while for 2D games, designers usually opt for \emph{Top Down}, \emph{Side View} or \emph{Isometric Perspective} cameras \cite{adams2014fundamentals}. 

\subsection{First Person Perspective}
\begin{figure}[h]
\includegraphics[width=8.75cm]{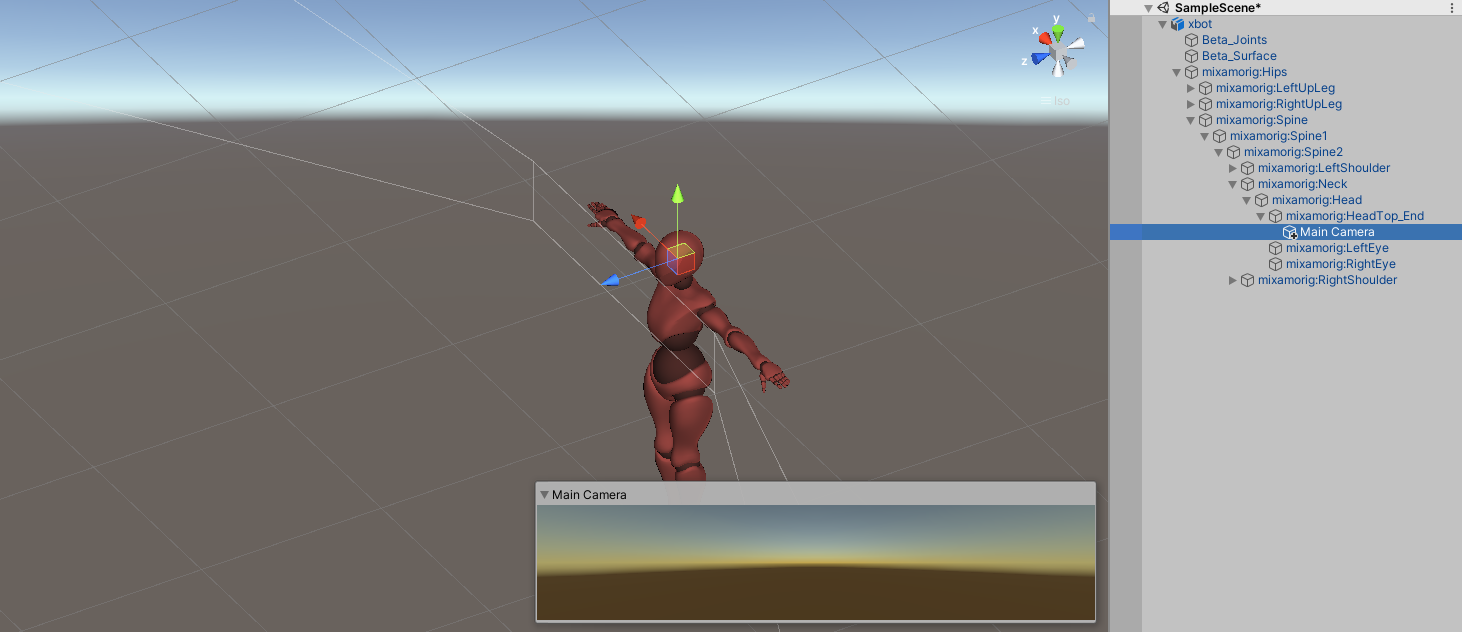}
\caption{Camera placement and orientation in the First Person Perspective model}
\label{fps_unity}
\end{figure}

\begin{figure}[h]
\includegraphics[width=8.75cm]{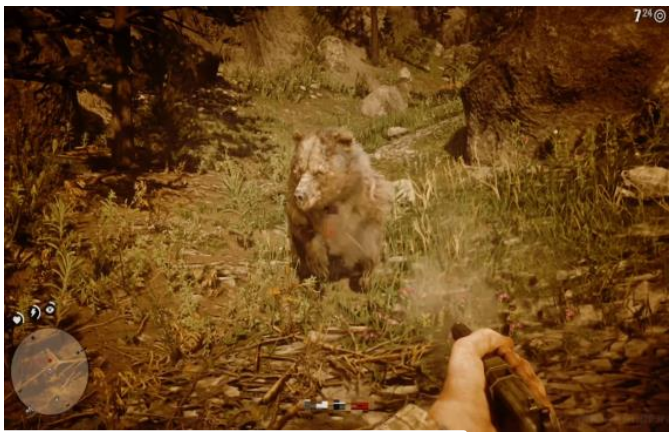}
\caption{Camera placement and orientation in Red Dead Redemption by Rockstar games}
\label{fps_rdr}
\end{figure}

In this model, the position and orientation of the camera correspond to where the player's character is and where it's looking at; thus, the virtual camera is usually positioned at the player's eye area and moves and rotates in the same manner as the character's head (Figure \ref{fps_unity}). This means that players cannot actually see their characters, although they may be able to see an object that the character carries (usually a weapon) or the character's hands (Figure \ref{fps_rdr}). As a result of this placement, players do not need (and are not able) to manipulate the camera view separately from their character \cite{adams2014fundamentals}.

\begin{figure}[h]
\includegraphics[width=8.75cm]{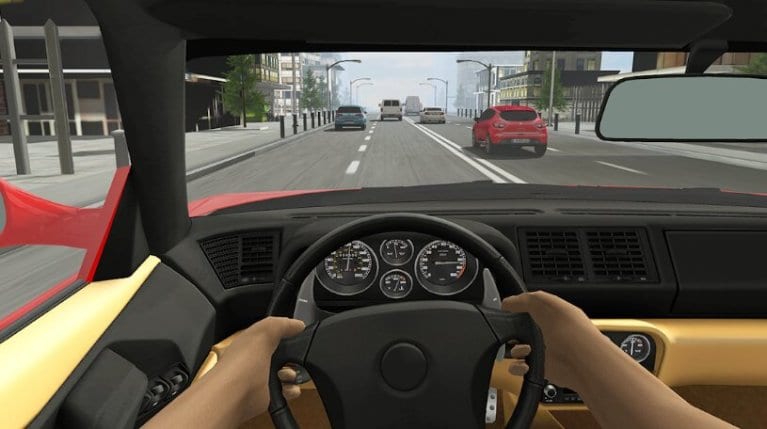}
\caption{Camera placement and orientation in Racing in Car 2 by Fast Free Games}
\label{fpsracing}
\end{figure}

A variant of this model can be found in racing games, where the camera view includes the road ahead of the player's vehicle and the car dashboard, but not the actual character itself (Figure \ref{fpsracing}). As a general rule, this camera model helps players identify and target approaching enemies, since it offers them a more natural and unobstructed view of their surroundings. In addition, the character's viewpoint is the same as the player's, so players do not need to compensate when aiming or using objects.

\subsection{Third Person Perspective}
\begin{figure}[h]
\includegraphics[width=8.75cm]{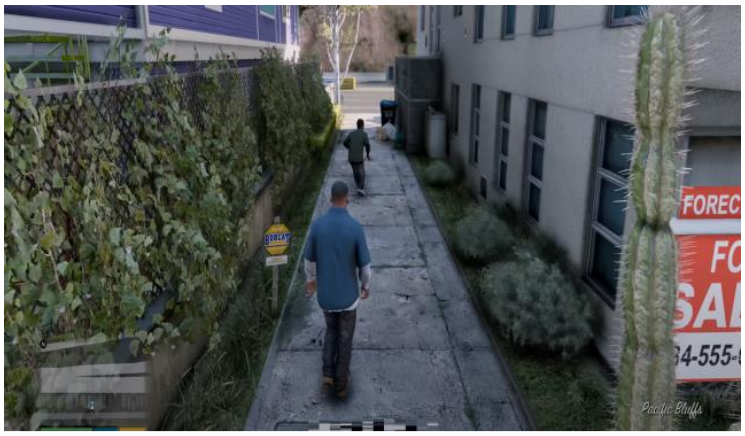}
\caption{Third Person Perspective camera in Grand Theft Auto V by Rockstar games}
\label{thirdperson}
\end{figure}

This is the camera model game designers prefer to use in action games, since it offers a view that includes both the player's character, as well as a part of its surroundings (Figure \ref{thirdperson}). In most implementations, the character's body is fully visible, the camera is placed behind it but close to its body and, again, follows its movement (but not necessarily its orientation, since that might induce motion sickness to the player). 

\begin{figure}[h]
\includegraphics[width=8.75cm]{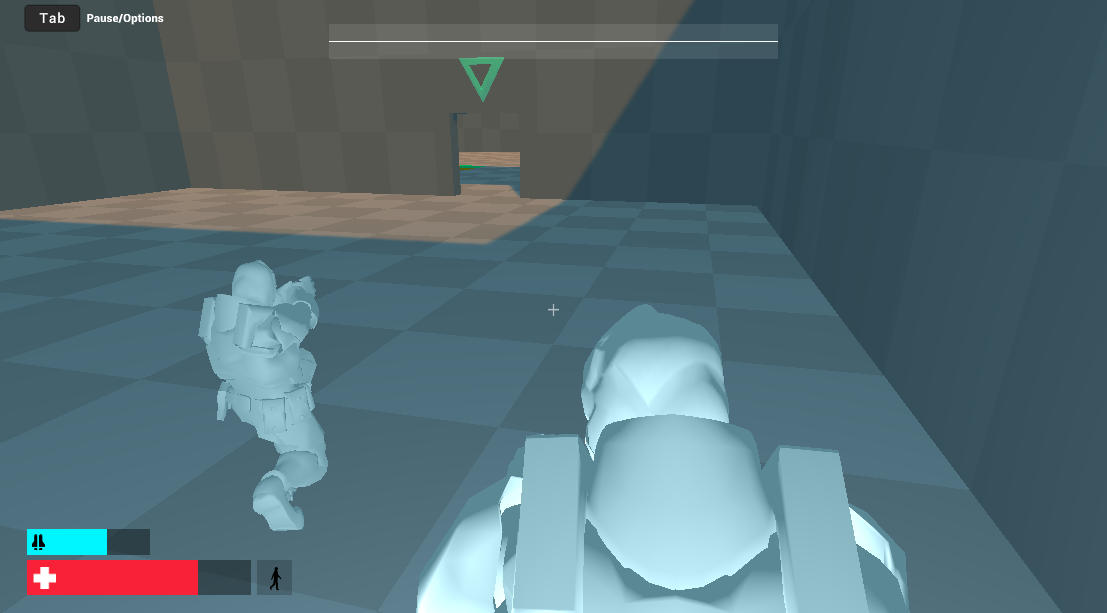}
\caption{Combination of a First-Person and Third-Person view}
\label{FPSvsTPS}
\end{figure}

Depending on the narrative and gamatography, this camera model may be divided into three sub-types, with the first using cameras in predefined points in the game level and activating the one which covers the area where the character is supposed to go next, the second following the character and moving along its path, and the third offering players the possibility to move its orientation. In the latter case, characters may run towards one direction and look in another, a mechanic which is extremely useful in fast-paced shooting or action games \cite{haigh2009real}. Overall, third person cameras enable players to see more of the action, as well as the body and movement of the game character (cf. Figure \ref{FPSvsTPS}) \cite{saltzman2004game}.

\subsection{Side View Perspective}
\begin{figure}[h]
\includegraphics[width=8.75cm]{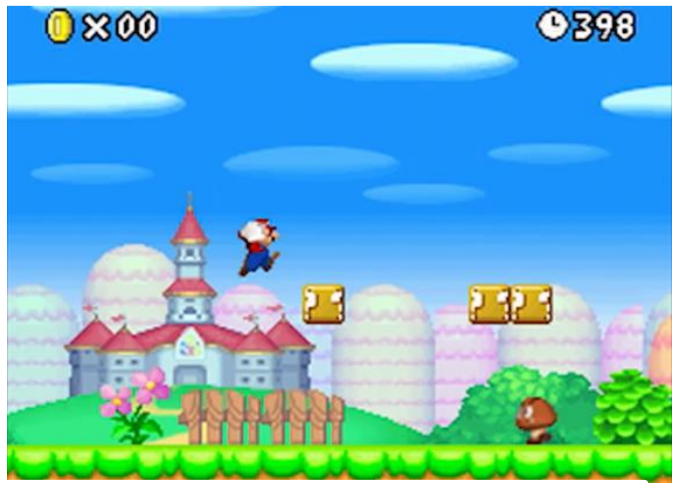}
\caption{Side View Perspective camera in New Super Mario Bros by Nintendo}
\label{sideview}
\end{figure}

With this camera model (Figure \ref{sideview}), the game world and game events are viewed from the side, with the perpendicular direction of the camera eliminating most of the depth of the scene. The player's character is typically on the one side of the view, as per usual cinematography conventions \cite{adams2014fundamentals}, making this setup more useful for games which are rich in narrative (including most fighting games, where players engage in one-to-one combat with enemies) or for linear game levels, where challenges come in sequences (e.g. platform games \cite{karpouzis2015platformer}) or in waves (e.g. side scroller games) \cite{saltzman2004game}.

\subsection{Top Down Perspective}
\begin{figure}[h]
\includegraphics[width=8.75cm]{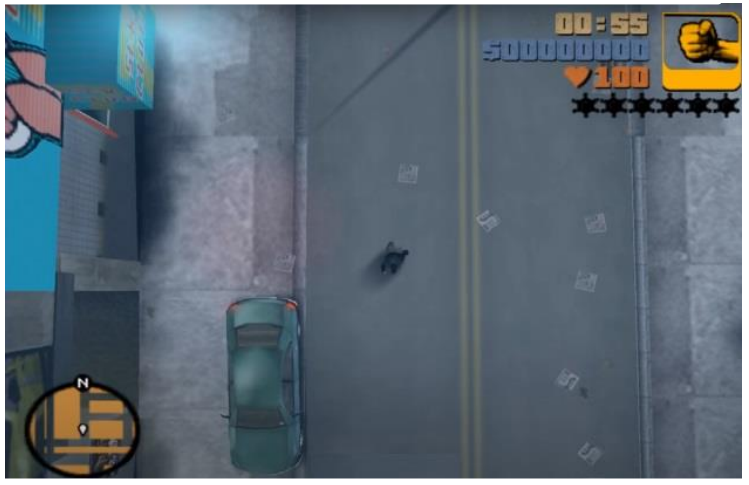}
\caption{Top Down Perspective camera in Grand Theft Auto III by DMA Design/Rockstar}
\label{topdown}
\end{figure}

Top-Down Perspective cameras offer an elevated view of the game world, offering players more spatial information about the characters surroundings (or the part of the game map the player chooses to inspect) \cite{galeos2020developing}. Cameras are placed above the game level and are oriented towards it (see Figure \ref{topdown}), making this model more suitable for 2D role-playing and real-time strategy games \cite{saltzman2004game}.

\subsection{Level design}
The purpose of game level design is to bring together aspects of narrative, spatial and aesthetic design and challenge-based engagement to make games fun, interesting and rewarding \cite{vargianniti2019effects}. Rouse \cite{rouse2004game} describes level design as ``elaborate game design'', which provides different results for each game, depending on the actual player experience we seek to evoke \cite{schell2008art}. In the vast majority of games, level design refers to the aesthetics and content of the game world or environment, the starting conditions of the level, any win conditions that players have to achieve to succeed in the level, and the interaction between the player character, the game world and the narrative. Some universal game design principles refer to all game genres, such as the game ``atmosphere'' or ambience, the pace of the game (whether it involves frantic action or thoughtful processes) and the tutorial levels offered to introduce players to the game, while genre-specific principles (such as camera model and placement) focus on game mechanics and challenges harmonized to the narrative of the game \cite{kotsia2016multimodal}. For instance, in video games where character exploration is important, the layout of the level constitutes a major factor of the player experience. In this context, designers usually opt for Open, Linear, Parallel, Ring, Hub or Network arrangements; in Linear levels, characters are placed in a game world without alternative hallways or branching, with challenges being set up in sequences and players moving only to adjacent parts of the level. As a result, this game design logic is more suitable for linear narratives with very little creative freedom left to players (\cite{karpouzis2016emotion}, \cite{karpouzis2021ai}).

\section{SpaceBall: Rob’s Adventure}
The game we created is called ``SpaceBall: Rob’s Adventure''. SpaceBall is an action-adventure game that starts as 3D environment, with players being able to view it as a 2D scroller. The game takes place in an abandoned, futuristic city with a robot attempting to collect various artefacts scattered around the game world. The robot has to overcome challenges of different kinds to succeed, such as traps, gaps and enemy characters. Depending on the actual part of the level and the challenges the character faces, players can change the active camera model to get a wider or more focused view of the game world. The game collects player behavior (i.e. actions in game world context and camera model switches) using Unity Analytics\footnote{Unity Analytics, https://unity.com/features/analytics}, a game analytics library included in the Unity3D game engine, used to transparently collect telematics information about the game world and its state, while preserving the privacy of the players.

\subsection{Camera choices in SpaceBall}
\begin{figure}[h]
\includegraphics[width=8.75cm]{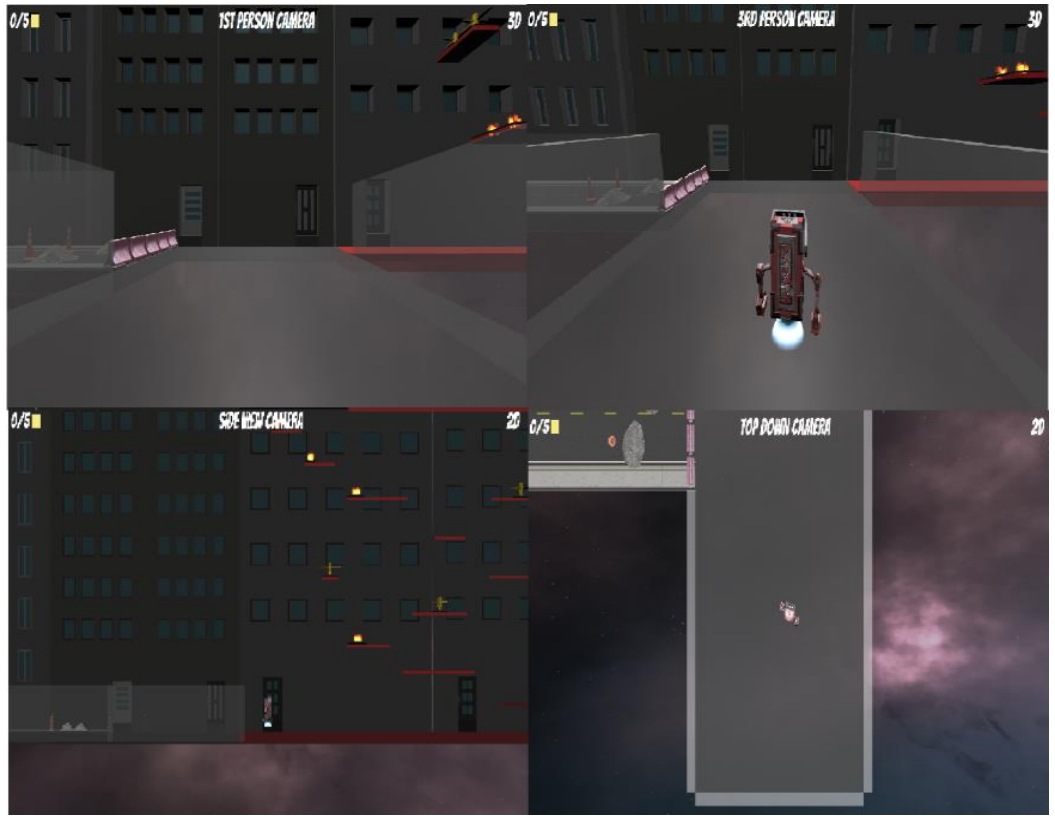}
\caption{Different camera models in SpaceBall: Rob’s Adventure}
\label{camera_models}
\end{figure}

Given the research question of SpaceBall, i.e. to induce, collect and analyze camera model choices given the game context, the game design process had to combine elements which do not normally co-exist in a single game. The first one was developing four possible camera models, two of which project the game world as a 3D space (First and Third Person Perspective), while the remaining two (Side view and Top-Down Orthographic) offer a 2D view of the character's surroundings. The idea here is that all camera models would create a uniform player experience and players would be given the option to change between the active camera model to find which fits them the most. The second element was the introduction of different scenarios corresponding to different subgenres, each of which was a better fit for some of the camera models, but still adhered to the general narrative of the game. This element had to undergo extensive play-testing and also required a lot of experimentation from the part of the players, so as to identify the camera model which provided the best view of the surroundings of the character and the challenge that the player had to overcome at any given point. Since Unity3D offers designers the possibility to create as many cameras as they want with each of them viewing the game world with a different camera model, the only issue here was that of choosing the appropriate mechanic to allow players to switch from one model to the next. We opted to develop this mechanic using the \emph{coroutine} functionality in Unity3D, so as to deactivate the previously active camera before switching to the next option (Figure \ref{camera_models}).

\subsection{Game scenarios}
\label{scenarios}
In SpaceBall: Rob’s Adventure, the scenarios/challenges that players face correspond to five discrete points in the level and include genre concepts from \emph{platform} games and \emph{shooters}. Depending on the actual scenario and the active camera, the game world can be viewed in either 2D or 3D and the challenge can be overcome with three of the four available cameras. Since we wanted  players to go through all challenges in the same sequence, we opted for a linear level design, i.e. without introducing any alternative pathways or teleports which would enable players to change the sequence of the challenges. Level design also includes rest periods between challenges; these serve as a safe space for experimentation with the different camera views and also prepare players for the next challenge to be faced, eliminating the need to display explicit instructional messages.

The first challenge consists of fixed platforms at specific heights with traps placed on them; players have to jump across the platforms and avoid the traps in order to succeed. For the second challenge, players have to navigate across narrow, crossing platforms and not fall down, while for the third one, they have to run across a terrace before the enemy catches up on them; here, the enemy is activated as soon as the character approaches it and players have the option to shoot the enemy and immobilize it for a few seconds. The fourth scenario consists of a jumping sequence across vertically moving platforms, leading up to the final scenario where players have to navigate a small urban park, locate and collect a pick-up item and complete their mission. Whenever players mistime their jumps or fail to avoid the enemies, they have to restart the particular challenge and their character is respawned at the previous rest point.

\begin{figure}[h]
\includegraphics[width=8.75cm]{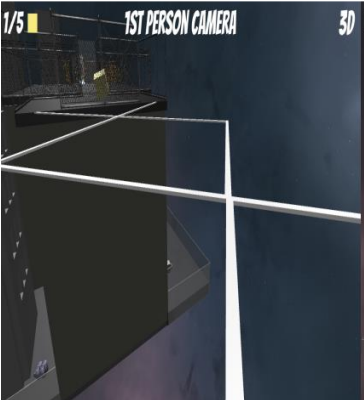}
\end{figure}

\begin{figure}[h]
\includegraphics[width=8.75cm]{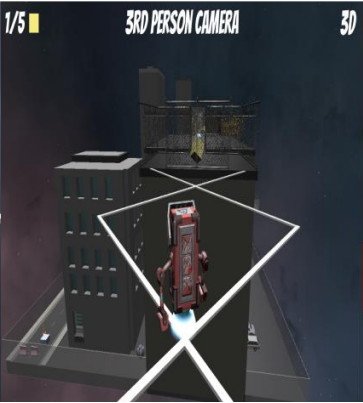}
\end{figure}

\begin{figure}[h]
\includegraphics[width=8.75cm]{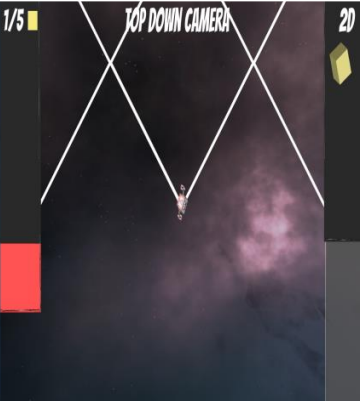}
\caption{A challenge where three different camera models can be used}
\label{three_camera_models}
\end{figure}

\section{Data collection, results and discussion}
\subsection{Data collection}
In order to answer our research question, we had to collect different kinds of information about the players (demographics and experience in gaming), as well as their game behavior and performance. We opted to avoid collecting post-game questionnaires about the latter, since that might compromise the integrity of the collected data, and used the Unity Analytics (UA) library, provided as part of the Unity game engine. This is a free service provided to game developers and offers the additional benefit of grouping collected data based on player ID (anonymized information); this is crucial, since we wanted to avoid collecting too much data from only a few players, reducing the scope of our results in the process. UA tracks two kinds of events, with the most usual being \emph{Standard events}, i.e. application events (usage of basic elements in the game UI), progression (player progress through the game), onboarding (the earliest interactions players have with the game, engagement (important actions related to social sharing and achievements) and monetization (any revenue-related events and in-game economies). In order to track the additional events necessary for our experiment, we used the \emph{Custom events} option and created one for each of the camera models and the five different scenarios/challenges; this was necessary, since Unity does not track context information explicitly for custom events, hence the need to relate different camera choices to each particular scenario. In addition to this, UA can only track 100 Standard events per hour and discards any extraneous information after that; during our tests, there were no actual cases where players sent more than 100 events, but this could easily be the case for longer games with more scenarios and options to choose from.

\subsection{Results and discussion}
SpaceBall was successfully completed by 30 players, all of them self-reported gamers and students of the SAE school of Game Programming. The figures below illustrate the camera of preference for each of the game scenarios (cf. Section \ref{scenarios}).

\begin{figure}[h]
\includegraphics[width=8.5cm]{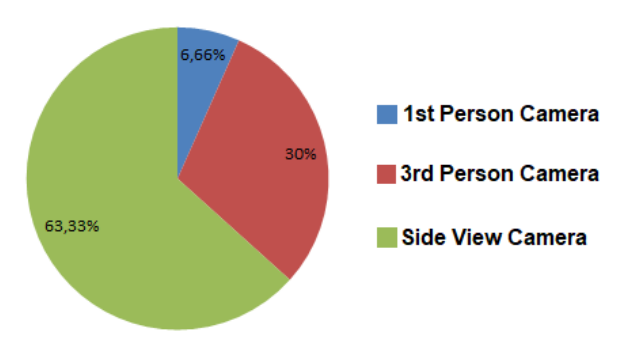}
\caption{Camera choices for Scenario 1: fixed platforms at specific heights}
\end{figure}

\begin{figure}[h]
\includegraphics[width=8.5cm]{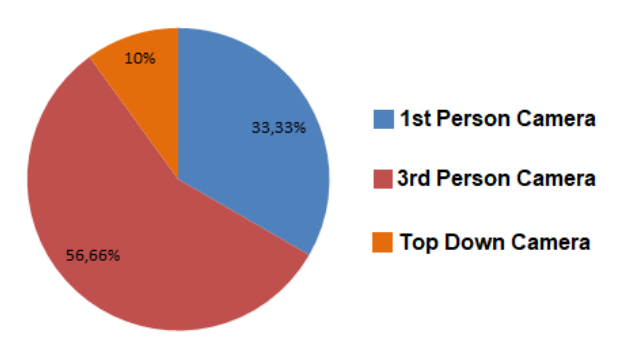}
\caption{Camera choices for Scenario 2: narrow, crossing platforms}
\end{figure}

\begin{figure}[h]
\includegraphics[width=8.5cm]{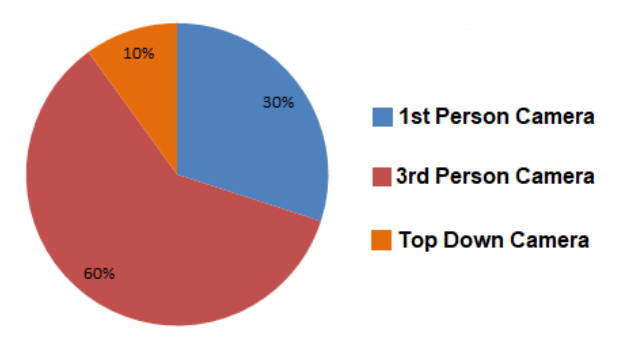}
\caption{Camera choices for Scenario 3: run across a terrace}
\end{figure}

\begin{figure}[h]
\includegraphics[width=8.5cm]{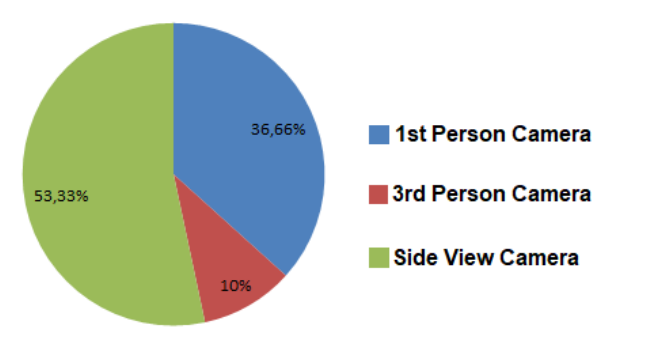}
\caption{Camera choices for Scenario 4: vertically moving platforms}
\end{figure}

\begin{figure}[h]
\includegraphics[width=8.5cm]{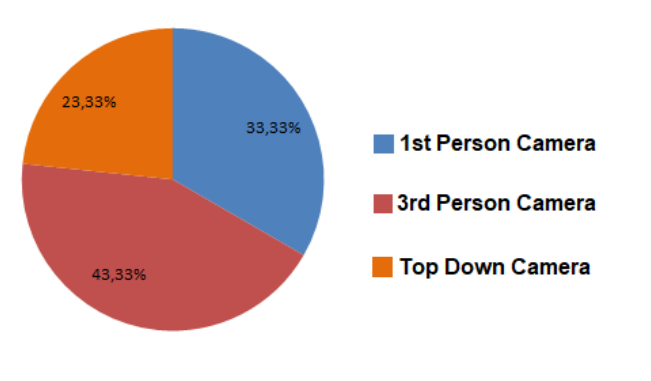}
\caption{Camera choices for Scenario 5: navigate a park, locate and collect a pick-up item}
\end{figure}

Overall, players opted to use 3D cameras in 68\% of the total successful challenges, and 2D cameras in the remaining 32\%. A 3rd Person camera was the most usual choice (40\%), followed by 1st Person (28\%) and Side View cameras (23,33\%). As mentioned in Section \ref{cameras}, each camera model presents a different view of the character's surroundings and challenges, and choosing the model that fits the most in each situation made perfect sense from the part of the players. Given the game genre and the game world, which mostly resembled a First-Person Shooter, the players' preference of cameras usually found in these genres was expected. In addition, challenges that involve jumping across platforms require richer spatial views in one plane, hence players opted to use a Side View camera in Scenario 4.

\section*{Acknowledgment}
This research has been co-financed by the European Union and Greek national funds through the Operational Program Competitiveness, Entrepreneurship and Innovation, under the call RESEARCH – CREATE – INNOVATE (project Mediludus, code: T2EDK-03049)

\bibliographystyle{IEEEtran}
\bibliography{MainDoc}

\end{document}